\documentstyle[eqsecnum,aps,multicol,bezier,amsmath,epsfig]{revtex}

\begin{document}

\draft

\title{Sub-Aging in a Domain Growth Model}
\author{L. Berthier}
\address{
Laboratoire de Physique -
ENS-Lyon and CNRS, 
F-69364, Lyon Cedex 07, France
}

\date{\today}

\maketitle 

\begin{abstract}
We study analytically the aging dynamics of the 
$O(n)$ model in the limit of $n \rightarrow \infty$,
with conserved and with non-conserved order parameter.
While in the non-conserved dynamics, the autocorrelation
function scales in the usual way $C(t,t_w) = {\cal C}(t/t_w)$, in 
the case of a conserved order parameter, `multiscaling' 
manifests itself in the form $C(t,t_w)={\cal C} (h(t)/h(t_w))$, with
a relaxation 
time growing more slowly than the age of the system (sub-aging),
and $h(t)$ a function growing faster
than any length scale of the problem.
In both cases, the effective temperature
associated to the violation of the fluctuation theorem
tends to infinity
in the asymptotic limit of large waiting times. 
\end{abstract}

\pacs{LPENSL-TH-02/2000}

\section{Introduction}

Aging of glassy systems is now well understood, at least 
from a qualitative point of view \cite{review}, and 
different approaches have been used to understand such a behaviour.
One of them is the interpretation of aging in terms of a
coarsening process.
The picture is the following: consider for instance an Ising ferromagnet,
which is quenched at time $t=0$ below its critical temperature.
When $t$ increases, two types of domains emerge, with up and down spins.
In the thermodynamic limit, equilibrium is never reached.   
At late times, domains have reached a typical
size $L(t)$. 
It is thus natural to assume scaling laws
for the different quantities of interest \cite{bray}. 
For instance, one can try the ansatz $S(\boldsymbol{k},t) \sim
L^d g(kL)$ for the structure factor (in a $d$-dimensional space), or
$C(t,t_w) \sim F(L(t)/L(t_w))$ 
for the two-time autocorrelation function, where
$g$ and $F$ are scaling functions.
The growth law $L(t)$ determines then
all the properties of the system.
As an example, the droplet model for spin glasses \cite{fisher}
assumes a logarithmic growth, leading 
to $C(t,t_w) = F(\ln t/\ln t_w)$.
If the growth law is given by 
$L(t) \sim t^\alpha$, like {\it e.g.} in a spinodal 
decomposition, one gets $C(t,t')=F(t/t_w)$.
This last behaviour is called `simple aging' and 
has been analytically shown to hold 
within various non-random models \cite{review,cude}.

Moreover, the above functional form
for the correlation function
is also found
analytically in some mean-field models of spin glasses, which 
give the general form 
for the correlation
functions in the aging regime   
$C(t,t_w)= {\cal C} ( \frac{h(t)}{h(t_w)} )$,
with $h$ and ${\cal C}$ two scaling functions \cite{review}
(valid in the two-time regime where
both times are large, but with $1<C<0$).
Although the notations are different, the functional form is
the same as in coarsening processes, and
it is then very natural to try to interpret the $h$-function as 
a relevant length scale for spin glasses, as was done for 
instance in ref.\cite{heiko}. 

From the experimental and numerical side,
it is found that a simple aging behaviour
describes the data well, in many different systems. 
This is interpreted by saying
that the relaxation time $t_r(t_w)$ of the system scales as 
the age $t_w$ of the sample: $t_r \sim t_w$.
However, a more subtle effect may appear, since
$t_r$ very often {\it grows more slowly than} $t_w$.
This effect has been called {\it sub-aging} \cite{vincent}.
In his pioneering experiments on polymer glasses, 
Struik \cite{struik} introduced the exponent $\mu$ from
 the relation $t_r \sim t_w^\mu$, with $\mu<1$.
Different values of $\mu$ have been reported:
Struik used $\mu \sim 0.9$, experiments in spin glasses $\mu \sim 0.97$
 \cite{vincent}, simulations of a structural 
glass were fitted using the value $\mu \sim 0.88$ \cite{jl}, 
and  recently, experiments on
a gel gave $\mu \sim 0.9$ \cite{luca}.
It can be checked (this point is discussed in detail in ref.\cite{vincent})
that the $\mu$-exponent 
is equivalent to the following choice 
of the $h$-function:
$
h(t) =\exp (\frac{1}{1-\mu} (\frac{t}{t_0})^{1-\mu} )
$.  
In accordance to what has been said above, this equivalence holds when 
$t_w \rightarrow \infty$ and $t - t_w \sim  t_w^{\mu}$.
Another function, 
the `enhanced power law' form 
$
h(t) = \exp ( \ln^a(t/t_0) )
$
with $a>1$,
has been phenomenologically introduced in
the context of spin glasses \cite{vincent},
and the value $a=2.2$ was used to fit experiments.
This in turn gives the relation $t_r \sim t_w/\ln^{a-1}(t_w)$,
valid in the regime $t_w \rightarrow \infty$ and 
$t-t_w \sim  t_w/\ln^{a-1}(t_w)$.

These choices are nonetheless not clearly motivated from a 
theoretical point of view, since
the mean-field spin glass models discussed above
only predict the existence of $h(t)$, and its 
analytical computation remains at present an open problem. 
In this context, simple models where $h$ can be computed
are much needed, but
there are only few examples where sub-aging appears. 
Very recently, a model exhibiting 
sub-aging has been proposed by Rinn {\it et al} \cite{bouchaud}, 
who studied a slight variation of Bouchaud's trap model for aging.
This has given a theoretical support to the use of an 
exponent $\mu$, even if its physical origin remains 
somewhat unclear.
A scaling approach to the diffusion of a point particle 
in a low dimensional space has been proposed in ref.\cite{laloux}, and
leads in some cases to a sub-aging which can be well
described by an enhanced power law.

We study in the present paper a model for coarsening
(the $O(n)$ model in the large-$n$ limit) which also
exhibits a sub-aging scaling in the autocorrelation function
when the order parameter is not conserved.
Its origin is the 
{\it simultaneous presence in the system of two 
different length scales,} 
whose consequence is the breakdown of the simple scaling laws
generally used in domain growth processes.
In particular, no $t/t_w$-scaling is found, and the relaxation time
grows as $t_r \sim t_w/\sqrt{\ln t_w}$ (sub-aging).
The autocorrelation is shown to be well
represented in the asymptotic regime 
by an enhanced power law with $a=3/2$,
{\it i.e.} $h(t)=\exp((\ln x)^{3/2})$.
Interestingly enough, $h(t)$ can not be interpreted in our
example as a length scale.
We do not want to argue that the model is
a realistic one for the aging of polymers or spin glasses,
but rather to give a possible physical explanation (the role 
of length scales \cite{footnote}) 
for the absence of the `naive' $t/t_w$-scaling, and exhibit a simple example
where the $h$-function can be computed and discussed in
terms of length scales, which has not been done so far.

\section{The $O(n)$ model}

This model is one of the few exactly solvable model
for coarsening.
It was first studied by Coniglio and Zannetti \cite{coza},
who computed the scaling properties
of the structure factor during the domain growth process.
They pointed out the presence of the two mentioned length scales,
and named `multiscaling' the breakdown of the usual $S(\boldsymbol{k},t) \sim
L^d g(kL)$.
Bray and Humayun have shown, however, that this multiscaling
was a peculiarity of the large-$n$ limit, and proved
that for a large but finite value of $n$, a `normal scaling' 
was recovered \cite{bray2}.
On an other hand, this `pathology' has been shown to appear
as a relevant preasymptotic effect in different 
coarsening models~\cite{caza}, like for instance the kinetic Ising model.

The model is defined through the Hamiltonian
\begin{equation}
H[\boldsymbol{\phi}]=\int d^d \boldsymbol{x} \left( 
\frac{1}{2} (\boldsymbol{\nabla}\boldsymbol{\phi})^2 + 
\frac{1}{4n} (n-\boldsymbol{\phi}^2)^2 \right),
\label{hamiltonian}
\end{equation}
where $\boldsymbol{\phi}(\boldsymbol{x},t)$ is a $n$-component vector field
in a $d$-dimensional space.
Two different dynamics may be associated to this model, depending on
whether or not the order parameter is conserved.
In the case of a non-conserved order parameter, the dynamics is given by
the so-called time dependent Ginzburg-Landau equation
\begin{equation}
\frac{\partial \boldsymbol{\phi(\boldsymbol{x},t)}}{\partial t} = -
\frac{\delta H}{\delta
\boldsymbol{\phi(\boldsymbol{x},t)}} + \boldsymbol{\eta}(\boldsymbol{x},t),
\label{tdgl}
\end{equation}
where $\boldsymbol{\eta}(\boldsymbol{x},t)$ is a random Gaussian variable
with mean zero and variance given by
$
\langle \boldsymbol{\eta}(\boldsymbol{x},t) 
\boldsymbol{\eta}(\boldsymbol{x'},t') \rangle = 2T \delta(t-t') \delta(
\boldsymbol{x} -\boldsymbol{x'}).
$

For conserved fields, we add $-\nabla^2$ in front of the r.h.s to get
the Cahn-Hilliard equation
\begin{equation}
\frac{\partial \boldsymbol{\phi(\boldsymbol{x},t)}}{\partial t} =
 \nabla^2 \left(
\frac{\delta H}{\delta
\boldsymbol{\phi(\boldsymbol{x},t)}} \right) 
+ \boldsymbol{\eta}(\boldsymbol{x},t),
\label{ch}
\end{equation}
where the variance of $\boldsymbol{\eta}(\boldsymbol{x},t)$ is
$
\langle \boldsymbol{\eta}(\boldsymbol{x},t) 
\boldsymbol{\eta}(\boldsymbol{x'},t') \rangle = - 2T \delta(t-t') 
 \nabla^2 \delta(
\boldsymbol{x} -\boldsymbol{x'}).
$

We shall see below that the limit $n \rightarrow \infty$ allows 
to solve the dynamics in both cases.
The key point that makes the model exactly soluble is that in the limit
of $n \rightarrow \infty$, the replacement
$
\boldsymbol{\phi}^2 / n \rightarrow 
\langle \phi^2 \rangle,
\label{trick}
$
where $\phi$ is one of the components of $\boldsymbol{\phi}$, can be made.
The two types of dynamics are now successively considered. 

\section{Non-conserved order parameter: simple aging}

The time dependent Ginzburg-Landau equation (\ref{tdgl})
associated to the Hamiltonian (\ref{hamiltonian}) is
\begin{equation}
\frac{\partial \boldsymbol{\phi}}{\partial t} =   
\nabla^2 \boldsymbol{\phi} + \boldsymbol{\phi} -\frac{1}{n}
(\boldsymbol{\phi}^2)\boldsymbol{\phi} + \boldsymbol{\eta},
\label{modela}
\end{equation}
where the dependence on space and time has been removed for clarity.
This differential equation is associated with random
initial conditions, in order to reproduce the quench experiment
described in the introduction, and
$\phi(\boldsymbol{x},0)$ is taken from a 
Gaussian distribution with zero mean and variance
$
\langle \phi(\boldsymbol{x},0) \phi(\boldsymbol{x'},0) \rangle = \Delta
\delta (\boldsymbol{x}-\boldsymbol{x'}).
$
From now on, we work at $T=0$. 
In the coarsening problem, temperature does not
play an essential role,
provided it is below the critical temperature.
The scaling 
regime can then be directly studied at $T=0$.
The review paper \cite{bray} provides a longer discussion of that point,
and we discuss below how our results may be (slightly)
changed by an non-zero temperature. 

The large-$n$ limit results in
the following equations
which have to be self-consistently solved:
\begin{equation}
\frac{\partial \phi}{\partial t} =   \nabla^2  
\phi + a(t)  \phi; \quad
a(t) =  1 -  \langle \boldsymbol{\phi}^2 \rangle.
\end{equation}
The solution is discussed in Refs.\cite{bray,coza}, and one finds
for the Fourier transform $\phi(\boldsymbol{k},t) = 
\int d^d \boldsymbol{x} \phi(\boldsymbol{x},t)
e^{ -i \boldsymbol{k} \cdot \boldsymbol{x} }$
\begin{equation}
\phi(\boldsymbol{k},t)=\phi(\boldsymbol{k},0)
e^{-k^2t}\left(\frac{t}{t_0} \right)^{d/4},
\end{equation}
where $t_0 \equiv \Delta^{2/d}/8\pi$.
It is now easy to compute the structure factor
\begin{equation}
S(\boldsymbol{k},t) \equiv \frac{1}{V} 
\langle \phi(\boldsymbol{k},t)
\phi(\boldsymbol{-k},t) \rangle 
= (8\pi t)^{d/2} e^{-2 k^2 t}. 
\label{simplescaling}
\end{equation}
We used $\langle \phi(\boldsymbol{k},0) \phi(\boldsymbol{-k},0)
\rangle = \Delta V$
from initial conditions.
The structure factor 
may be written as $S(\boldsymbol{k},t) = L^d g(kL)$, with
$L(t) = t^{1/2}$ and $g(x)= (8 \pi)^{d/2} \exp (-2x^2)$, demonstrating 
the validity of the scaling hypothesis in that case.

The autocorrelation function is defined as
\begin{equation}
C(t,t_w) \equiv \frac{1}{V} \int {d}^d \boldsymbol{x} \langle 
\phi(\boldsymbol{x},t) \phi(\boldsymbol{x},t_w) 
\rangle 
=  \frac{1}{V}\int \frac{d^d \boldsymbol{k}}{(2\pi)^d} 
\langle \phi(\boldsymbol{k},t) 
\phi(\boldsymbol{-k},t_w)
\rangle
\end{equation}
and may be easily computed:
\begin{equation}
C(t,t_w)=\left[  \frac{2 \sqrt{tt_w}}{t+t_w} \right]^{d/2}.
\label{c1}
\end{equation}
Defining the scaling variable $\lambda_1 \equiv t/t_w$, $C(t,t_w)$ can 
be rewritten
\begin{equation}
C(t,t_w)= F_1(\lambda_1);\quad 
F_1(x) \equiv \left[  \frac{2 \sqrt{x}}{1+x} \right]^{d/2} .
\label{scaling}
\end{equation}
This last equation means that the autocorrelation
function exhibits a simple aging behaviour.
We have then illustrated on a concrete
model the scaling approach to domain growth described
in the introduction. 
We shall see in the next section the
differences arising when sub-aging is present.

Let us note here that a finite 
temperature does not affect the above discussion, 
since it simply introduces a short-time relaxation in the
correlation function, that does not depend on the waiting time
$t_w$ and corresponds to an {\it equilibrium relaxation inside the
growing domains}.
The long-time relaxation we are interested in,
and which corresponds to
the growth of the domains themselves 
is still described by (\ref{scaling}).  

\section{Conserved order parameter: sub-aging}

The Cahn-Hilliard equation (\ref{ch})
associated to the Hamiltonian (\ref{hamiltonian})
is given by (still at $T=0$)
\begin{equation}
\frac{\partial \boldsymbol{\phi}}{\partial t} = -\nabla^2  
\left[ \nabla^2 \boldsymbol{\phi} + \boldsymbol{\phi} -\frac{1}{n}
(\boldsymbol{\phi}^2)\boldsymbol{\phi} \right],
\label{modelb}
\end{equation}
and is solved following 
the same steps as previously, 
leading to \cite{bray,coza}:
\begin{equation}
\phi(\boldsymbol{k},t) = \phi(\boldsymbol{k},0) \exp \left( -k^4t+k^2 
\sqrt{\frac{dt}{2}\ln (\frac{t}{t_0})} \right),
\label{solution}
\end{equation}
with $t_0 \equiv \Delta^{4/d}/(16 \pi)^2$.
The structure factor reads in that case
\begin{equation}
S(\boldsymbol{k},t) \sim  
 \left[ L_1(t)^{d} \right]^{f( k L_2(t))},
\label{multiscaling}
\end{equation}
where $f(x) \equiv  2x^2-x^4$. 
In this expression, two characteristic length scales have been
defined:
$
L_1(t) \equiv t^{1/4}
$, and
$L_2(t) \equiv ( \frac{8t}
{d\ln (t/t_0)} )^{1/4}$.
In the standard scaling form, $S(\boldsymbol{k},t) \sim
L^d g(kL)$, the structure factor 
varies as $L^d$ with a prefactor depending on the
scaling variable $kL$, whereas
for the multiscaling form (\ref{multiscaling}),
$S$ varies as $L_1^\alpha$, with an exponent $\alpha$ which depends
continuously on the scaling variable $kL_2$. 
The two scalings are thus completely different.

Coniglio and Zannetti \cite{coza} have
interpreted this multiscaling in 
terms of domains composed of sub-domains, each sub-domain 
growing at a different rate.
The initial motivation for the present work was indeed 
to investigate the possible existence of
a `hierarchy' of time scales, similar to the one
found in mean-field spin glass models
(`ultrametricity in time') 
\cite{review,bouchaud,cuku}.
A different effect arises instead. 
Using eq.(\ref{solution}), one easily
gets for the autocorrelation function 
\begin{equation}
C(t,t_w) \sim \frac{1}{(t+t_w)^{d/4}} \exp \left( \frac{d}{8} \frac{
\left(  \sqrt{ t \ln (t/t_0)} 
+ \sqrt{ t_w \ln (t_w/t_0)} \right)^2 }{
t+t_w} \right).
\label{correlation}
\end{equation}
It is obvious from this expression that $C(t,t_w)$ cannot be
written as a function of $t/t_w$ only.
The physical key ingredient for the absence of the usual scaling
is the presence of two different length scales in the system.
 
We prove now analytically 
that eq.(\ref{correlation}) implies sub-aging. 
It has to be remarked first that when the time difference
$\tau \equiv t-t_w$  is equal to $t_w$, one has
\begin{equation}
C(t_w + t_w,t_w) \underset{t_w \rightarrow \infty}{\sim} \frac{1}
{t_w^{(3-2\sqrt{2})d/24}} \rightarrow 0.
\end{equation}
In the asymptotic limit of large waiting times,
the relaxation of $C(t,t_w)$ is complete in times $\tau \ll t_w$.
In that regime, one can show that
\begin{equation}
C(t,t_w) \underset{\tau \ll t_w}{\sim} \exp \left( - 
\frac{d \ln t_w}{64} \left( \frac{\tau}{t_w} \right)^2 \right).
\end{equation} 
Defining the scaling variable
$\lambda_2 \equiv \tau \sqrt{\ln t_w} / t_w$,
eq.(\ref{correlation}) can finally be rewritten
\begin{equation}
C(t,t_w) \sim F_2(\lambda_2); \quad  F_2(x) \equiv \exp \left(
- \frac{d x^2}{64} \right).
\end{equation}
The relaxation time grows hence as $t_r \sim t_w/\sqrt{\ln t_w}$, 
{\it i.e.} more slowly than $t_w$: {\it this is a sub-aging behaviour}.
It is moreover possible to compute the function $h(t)$ discussed 
in the introduction.
The scaling form $C(t,t_w)= {\cal C} \left( \frac{h(t)}{h(t_w)} \right)$
should be valid in the two-time regime where
both times are large, but with a non-zero
value of the correlation function.
In the present case, this regime
is characterized by
\begin{equation}
t_w \rightarrow \infty, \quad \tau \sim \frac{t_w}{\sqrt{\ln t_w}}.
\label{regime}
\end{equation}
We have seen that a natural choice for $h(t)$ would be $L_1(t)$ or 
$L_2(t)$, {\it i.e.} a length scale, since it is a common interpretation.
{\it This does not work}, and a 
more complicated form has to be found.
It is straightforward to realize that a possible choice is
an enhanced power law:
\begin{equation}
{\cal C}(x) = \exp \left( - \frac{d}{288} \ln^2(x) \right); \quad
h(t) =  \exp \left( (\ln t)^{3/2} \right).
\end{equation}
The function $h$ is neither $L_1$ nor $L_2$, but a combination of the two,
and therefore does not have a direct physical interpretation:
$
h(t) \sim \exp \left(  \left( L_1 /L_2  \right)^6 \right).
$

\section{Response functions: infinite effective temperatures}

It is also relevant to study the response functions for aging 
systems, since it is a major prediction of the dynamical mean-field
theory for spin glasses that interesting informations are
encoded in the susceptibilities \cite{review,teff}.
Up to now, we have studied aging in
the two-time correlation functions $C(t,t_w)$.
In glassy systems, aging is also found in the related response 
functions $R(t,t_w)$, associated with a 
breakdown of the fluctuation dissipation theorem which at
equilibrium would be $T R(t,t_w) = \partial C(t,t_w)/\partial t_w$.
This is taken into account by introducing an
effective temperature $T_{\text{eff}}$ through~\cite{teff}
\begin{equation} 
T_{\text{eff}} (q) = \lim_{t_w \rightarrow \infty} 
\frac{\frac{\partial C(t,t_w)}{\partial tw}}{R(t,t_w)}
\Bigg{\vert}_{C(t,t_w)=q} .
\end{equation}

In coarsening systems, however, response functions have 
been shown numerically and analytically to be weak, in the
sense that $T_{\text{eff}} \rightarrow \infty$
\cite{barrat,bebaku}.
This property has been related to the decreasing density of 
topological defects (domain walls) during the coarsening.
In the case of the $O(n)$ model, no topological defects are
present if $n>d$, which is naturally the case in the large-$n$ limit.
We compute then $R(t,t_w)$ in the both cases studied 
above to obtain $T_{\text{eff}}$.
We refer the reader to ref.\cite{bebaku} for the method,
since we follow exactly the same steps.
We get the two following expressions:
\begin{equation}
R (t,t_w) \sim 
\left( \frac{t}{t_w} \right)^{d/4}   \left( \frac{1}{t-t_w} \right)^{d/2},
\label{r1}
\end{equation}
in the non-conserved case, and
\begin{equation}
R (t,t_w) \sim
 \frac{1}{(t-t_w)^{(d+2)/4}} \exp \left( \frac{d}{8} \frac{
\left(  \sqrt{ t \ln t} 
- \sqrt{ t_w \ln t_w} \right)^2 }{
t-t_w} \right),
\label{r2}
\end{equation}
in the conserved case (we dropped out all numerical constants).
Combining eqs.(\ref{c1},\ref{correlation},\ref{r1},\ref{r2}),
it is easy to show that for the non-conserved and the conserved case
successively, one has:
\begin{equation}
T_{\text{eff}} (q) \sim  \lim_{t_w \rightarrow \infty} t_w^{d/2-1},\quad
T_{\text{eff}} (q) \sim  \lim_{t_w \rightarrow \infty} \frac{t_w^{(d-2)/4}}
{ (\ln t_w)^{(d+2)/8} \exp(\sqrt{\ln t_w})}.
\end{equation}
This holds for $0<q<1$, and shows that for $d>2$,
although there is no interpretation here in terms of defects,
the effective temperature is infinite,
as has been found so far in all
domain growth processes \cite{cude,barrat,bebaku}.

We studied in this letter the aging dynamics of the $O(n)$ model in
the large-$n$ limit. We showed that when the order parameter is not
conserved, standard scaling laws hold, leading to a simple
aging behaviour.
We investigated the more interesting case of a conserved dynamics,
and were able to show that the multiscaling observed in
the structure factor does not imply 
a hierarchy of time scales (`ultrametricity
in time' \cite{cuku}).
Rather, the relaxation takes place in a time scale which is shorter
than the waiting time, $t_r \sim t_w/\ln^{a-1}(t_w)$ with $a=3/2$,
the correlation function being well represented in that
regime by $C(t,t_w)={\cal C} (h(t)/h(t_w))$, where $h$ 
is an enhanced power law $h(t)=\exp(\ln^a(t))$.
This simple example exhibits then a very rich aging behaviour,
whose origin is the presence of two different length scales during
the coarsening process. 
It shows also
that the interpretation of $h(t)$ as a
length scale may in some cases be misleading. 
The enhanced exponential form 
that has been successfully used to fit 
spin glass experiments
arises naturally 
from our computation.
It implies that the relaxation time
scales as $t_r \sim t_w/\ln^{a-1}(t_w)$, which could 
hardly be experimentally distinguishable 
from a power law $t_r \sim t_w^{\mu}$, when
$\mu$ is very near to one, as it is in spin glasses.

\acknowledgments
I sincerely thank {\sc J. Kurchan} who suggested and followed this work,
{\sc J.-L. Barrat} and {\sc J.-Ph. Bouchaud} for their 
interest and encouragements,
{\sc L. F. Cugliandolo} and {\sc M. Sellitto} for their help during
the preparation of the manuscript.


\begin{thebibliography}{0}

\bibitem{review}
Bouchaud J.-Ph., Cugliandolo L. F., Kurchan J. and 
M\'ezard M.,
{\it Spin Glasses and Random Fields}, edited by 
Young A. P.
(World Scientific, Singapore)
1998;
Bouchaud J.-Ph., 
Lectures notes for the ``{\it Soft and fragile matter}''
Summer School, St Andrews 1999, preprint cond-mat/9910387.

\bibitem{bray} 
Bray A. J.,
Adv. in Phys. {\bf 43} (1994) 357.

\bibitem{fisher} 
Fisher D. S.,
Physica D {\bf 107} (1997) 204; 
Fisher D. S. and Huse D. A., 
Phys. Rev. B {\bf 38} (1998) 373. 

\bibitem{cude}
Shukla P. and Singh S.,
Phys. Rev. B {\bf 23} (1981) 4661; 
Ciuchi S. and de Pasquale F.,
Nucl. Phys. B {\bf 300} (1988) 31;
Cugliandolo L. F. and Dean D. S., 
J. Phys. A {\bf  28} (1995) 4213; 
Godr\`eche C. and Luck J.-M.,
J. Phys. A {\bf 30} (1997) 6245;
Barrat A., Burioni R. and M\'ezard M.,
J. Phys. A {\bf 29} (1996) 1331;
Cugliandolo L. F., Kurchan J. and Parisi G.,
J. Phys. A {\bf 27} (1994) 5749;
Zippold W., Kuehn R. and Horner H.,
preprint cond-mat/9904329.

\bibitem{heiko} 
Kisker J., Santen L., Schreckenberg M.
and Rieger H.,
Phys. Rev. B {\bf 53} (1996) 6418.

\bibitem{vincent} 
Vincent E., Hammann J., Ocio M., Bouchaud J.-Ph. and Cugliandolo L. F.,
{\it Complex behaviour of glassy systems},
edited by Rubi M.
(Springer Verlag, Berlin) 1997.

\bibitem{struik} 
Struik L. C. E.,
{\it Physical aging in amorphous
polymers and other materials}
(Elsevier, Amsterdam) 1978.

\bibitem{jl} 
Kob W.and Barrat J.-L.,
Phys. Rev. Lett. {\bf 78} (1997) 4581.

\bibitem{luca} 
Cippelletti L., Manley S., Ball R. C.
and Weitz D. A.,
Phys. Rev. Lett. {\bf 84} (2000) 2275.

\bibitem{bouchaud} 
Rinn B., Maass P. and Bouchaud J.Ph.,
preprint cond-mat/0001161.

\bibitem{laloux}
Laloux L. and Le Doussal P.,
Phys. Rev. E {\bf 57} (1998) 6296.

\bibitem{footnote}
Raising the problem of finite size effects in spin glass
experiments, a possible explanation for sub-aging has been given in: 
Joh Y. G., Orbach R., Wood G. G., Hammann J.
and  Vincent E.,
preprint cond-mat/0002040;
Bouchaud J.-Ph., Vincent E. and Hammann,
J. Phys. I (France) {\bf 4} (1994) 139.

\bibitem{coza} 
Coniglio A. and Zannetti M.,
Europhys.  Lett. {\bf 10} (1989) 575.

\bibitem{bray2} 
Bray A. J. \and Humayun K.,
Phys. Rev. Lett. {\bf 68} (1992) 1559. 

\bibitem{caza} 
Castellano C. and Zannetti M.
cond-mat/9906094;
Castellano C. and Zannetti M.,
Phys. Rev. E {\bf 58} (1998) 5410.

\bibitem{cuku}
Cugliandolo L. F. and Kurchan J.,
J. Phys. A {\bf 27} (1994) 5749.

\bibitem{teff}
Cugliandolo L. F., Kurchan J. and Peliti L.,
Phys. Rev. E {\bf 55} (1997) 3898.

\bibitem{barrat} 
Barrat A.,
Phys. Rev. E {\bf 57} (1998) 3629.

\bibitem{bebaku} 
Berthier L., Barrat J.-L. and Kurchan J.,
Eur. Phys. J. B {\bf 11} (1999) 635.

\end{thebibliography}
\end{document}